\documentclass[12pt]{iopart}
\usepackage{bm,amsfonts,amssymb, graphicx}
\usepackage{epstopdf}
\begin{document}
\paper{Incoherent excitation of few-level multiatom ensembles}
\author{Baranidharan Mohan$^{1}$ and Mihai A. Macovei$^{1,2}$ }
\address{$^{1}$Max-Planck-Institut f\"{u}r Kernphysik, Saupfercheckweg 1, D-69117 Heidelberg, Germany}
\address{$^{2}$Institute of Applied Physics, Academiei str. 5, MD-2028 Chi\c{s}in\u{a}u, Moldova}
\eads{\mailto{ph1100833@physics.iitd.ac.in}}
\eads{\mailto{mihai.macovei@mpi-hd.mpg.de}}
\date{\today}
\begin{abstract}
We investigate the collective interaction of a multi-atom three-level ladder system with an
environmental incoherent reservoir. The exact solution of the master equation that describes
such a system of atoms emitting distinguishable or indistinguishable photons on interaction
with an incoherent reservoir is obtained. The quantum decay interference effects have no
influences over steady-state distribution of population among the energy levels. However, 
the photon statistics of the collectively emitted photons are different in this case. 
In particular, the normalized second-order photon correlation function shows a minimum in 
its value for the photons generated on lower atomic transition in a system with orthogonal 
dipoles. The minimum is due to the steady-state amplified spontaneous emission by the 
incoherent applied fields.
\end{abstract}
\submitto{\JPB}
\pacs{42.50.Fx, 42.50.Lc, 42.50.Ar}
\maketitle

\section{Introduction}
The collective behavior of quantum systems on interaction with a common radiation field was
first studied by Dicke~\cite{dik_et}. Since then, the studies regarding the collective
interactions of an ensemble of few-level emitters with and via an environmental reservoir and
pumped by external coherent or incoherent sources of electromagnetic light have been carried 
out intensively~\cite{dik_et,rev,has, ag_et, bog, kn_et}. The investigation of such interactions has
interested the scientific community because of the interesting effects and the possible application
in the field of quantum communication and quantum information. The manipulation of collective
fluorescence of an atomic sample via classical coherent fields and a heat bath has been shown in
Ref.~\cite{mek2}. Further, it has been shown that  the photon scattering by a collection of
few-level emitters in incoherent environments leads to violation of the Cauchy-Schwarz
inequality~\cite{mek}. Thus, it can be realized that, in contrary to general intuition, quantum
features can be obtained from the interaction of quantum systems with a classical electromagnetic
field (EMF) reservoir. This opens up a lot of possible applications in quantum information science,
for example, entanglement between two arbitrary qubits has been shown to be generated when they
interact with a common thermal bath \cite{ent1, ent2, ent3, ent4}. Recently, other interesting
developments like, stationary entanglement at high temperatures for two coupled, parametrically
driven, dissipative harmonic oscillators \cite{ent5} and room-temperature steady-state
optomechanical entanglement on a chip \cite{ent6}, have been shown. Disentanglement versus
decoherence of two qubits in thermal noise was investigated as well \cite{disent}.

With such a motivation, in this paper, we investigate the quantum behaviors of a collection of
three-level ladder emitters surrounded by an incoherent reservoir. We find that the steady-state
distribution of radiators on the energy levels are not affected by the presence of the cross-damping
terms caused by the interference of transition amplitudes. The collective effects drive the system
into a final thermal steady-state which is other than Boltzmann equilibrium distribution. The
photon statistics changes from super-Poissonian to sub-Poissonian depending on the number of
atoms in the sample, temperature and the mutual orientation of the induced dipole-moments.
In particular, we analyzed the steady state intensity and the normalized second-order correlation
function for the light generated on the lower atomic transition. We found that there occurs a
maximum in the steady-state intensity for moderately large atomic samples with orthogonal
transition dipoles. The physics behind such a behavior is that the steady-state intensity of
the emitted photons for the lower transition has a maximum due to the amplified spontaneous
emission by the incoherent field. Correspondingly, a minimum is observed in the normalized
second-order correlation function, resulting in the emission of quasi-coherent light. When the
orientation of the transition dipoles are near parallel, the dip is not observed and the photon
statistics is similar to that of a two-level ensemble.

The paper is organized as follows. In Section 2, we consider the system of interest and obtain the
exact steady-state solution of the master equation that describes the system. Using the solution,
we arrive at the distribution of the emitters on the atomic states. Section 3 investigates the photon
statistics of the spontaneously emitted photons as function of number of atoms, bath characteristics
and orientation of atomic dipoles. The results are summarized in Section 4.

\section{Master Equation and its exact steady state solution}
The basic element of our investigation is a sample of $N$ identical non-overlapping three-level
ladder emitters that interacts with an environmental incoherent reservoir like thermal bath or
broad-band incoherent lasers. The radiating atoms are located within a volume with linear dimension
less compared to the relevant emission wavelengths $\{\lambda_{12},\lambda_{23} \} $ (Dicke model). However, the obtained results apply to extended atomic samples as well where one atomic dimension is much larger than the relevant emission wavelength \cite{Eberly}.
The incoherent reservoir induces transitions between the atomic levels with rates proportional to
the mean incoherent photon number at corresponding atomic transitions. The excited atomic level
$| 1 \rangle$ $(|2\rangle)$ spontaneously decays to the state $|2\rangle$ $(|3\rangle)$ due to
the zero point fluctuation of the electromagnetic field, with a decay rate $2\gamma_1 (2\gamma_2)$.

In the usual mean field, Born-Markov, dipole and rotating wave approximations, the interaction of
the atomic sample with the surrounding incoherent bath is described by the master equation \cite{rev}:
\begin{eqnarray}
 &{}&\frac{d}{dt}{\rho}(t) + i[\omega_{12}S_{11}-\omega_{23}S_{33},\rho] = -(1+\bar{n}_1)[S_{12},(\gamma_1S_{21}
 + \gamma_{21}S_{32})\rho] \nonumber \\
&-& (1+\bar{n}_2)[S_{23},(\gamma_2S_{32}+\gamma_{12}S_{21})\rho]
-\bar{n}_1[S_{21},(\gamma_1S_{12}+\gamma_{21}S_{23})\rho] \nonumber \\
&-& \bar{n}_2[S_{32},(\gamma_2S_{23}+\gamma_{12}S_{12})\rho] + H.c.,   \label{master}
\end{eqnarray}
where the collective atomic operators
$S_{\alpha\beta} = \Sigma_{j=1}^N S_{\alpha\beta}^{(j)} =
\Sigma_{j=1}^N |\alpha\rangle_{jj}\langle\beta| \equiv |\alpha\rangle\langle\beta|$
describe the transitions between $|\beta\rangle$ and $|\alpha\rangle$ for $\alpha\neq\beta$
and populations for $\alpha = \beta$ and obey the commutation relation
$[S_{\alpha\beta},S_{\alpha^\prime\beta^\prime}] = \delta_{\beta\alpha^\prime}S_{\alpha\beta^\prime} 
- \delta_{\beta^\prime\alpha}S_{\alpha^\prime\beta}$. Here, $\bar n_i$ are the mean photon numbers
that represent the intensity of incoherent pumping. For thermal bath, the mean thermal photon
number is given by,
$$
 \bar{n}_i = \frac{1}{\exp(\beta\hbar\omega_{i,i+1})-1},
$$
where, $\omega_{i,i+1} = \omega_i - \omega_{i+1}, \{i=1,2 \}$ and $\beta = (k_BT)^{-1}$
where $k_B$ is the Boltzmann constant and {T} is the temperature of the bath.
For incoherent pumping,
$$
 \bar{n}_i = \frac{R_{i,i+1}d^2_{i,i+1}}{\gamma_i\hbar^2},
$$
where $R_{i,i+1}$ describes the strength of the incoherent pumping.
Furthermore, $2\gamma_{1(2)}$ = $4d^{2}_{12(23)}\omega^{3}_{12(23)}/(3\hbar c^{3})$ is the
single- atom natural line width. $\gamma _{12} = \gamma _{2}\frac{|\vec{d}_{12}|}{|\vec{d}_{23}|}\cos{\theta}$
and $ \gamma_{21} = \gamma_{1}\frac{|\vec{d}_{23}|}{|\vec{d}_{12}|}\cos{\theta}$, with
$\theta $ being the angle between the dipole moments $\vec{d}_{12}$ and $ \vec{d}_{23}$,
describe the interference (cross-damping) effects among the atomic transitions
$|1\rangle \leftrightarrow |2\rangle $ and $|2\rangle \leftrightarrow |3\rangle$, and
cannot be neglected for non-orthogonal dipole moments if
$\Delta$ = $|\omega_{12}-\omega_{23}| \le \Gamma_{eff} = N\gamma_{1(2)}[1+\bar n_{1(2)}]$.

The steady-state solution of the master equation, Eq.~(\ref{master}) is given by the relation:
\begin{equation}
 \rho_s = Z^{-1}e^{-\xi_1 S_{11}}e^{-\xi_3 S_{33}},     \label{master_solution}
\end{equation}
where for $\gamma_{12} = \gamma_{21} = 0 $ (i.e. when $\vec{d_{12}} \perp \vec{d_{23}}$),
$$
\xi_1 = \ln\bigg[\frac{1+\bar n _1}{\bar n _1}\bigg],\hspace{0.3cm}
\xi_3 = \ln\bigg[\frac{\bar n _2}{1+\bar n _2}\bigg]
$$
while for $\bar n _1 = \bar n _ 2 \equiv \bar n $ and
$\gamma_{12} = \gamma_{21} \neq 0 ,$
$$
\xi_1 = - \xi_3 \equiv \xi = \ln\bigg[\frac{1+\bar n}{\bar n}\bigg].
$$
Here $Z$ is chosen such that ${\rm Tr\{\rho_{s}\}=1}$. It is interesting to emphasize that the exact
solution, i.e. Eq.~(\ref{master_solution}), has a diagonal form while the master equation itself
Eq.~(\ref{master}) has non-diagonal terms which arise due to the two transitions
$ |1\rangle \rightarrow |2\rangle $ and $ |2\rangle \rightarrow | 3\rangle $ coupling with the same
electromagnetic field modes. The solution of the master equation (\ref{master}) was obtained by the
direct substitution of Eq.~(\ref{master_solution}) in the steady-state form of Eq.~(\ref{master}), 
noting that: $e^{\xi_{1}S_{11}}S_{21}e^{-\xi_{1}S_{11}}=S_{21}e^{-\xi_{1}}$ and 
$e^{\xi_{3}S_{33}}S_{32}e^{-\xi_{3}S_{33}}=S_{32}e^{\xi_{3}}$.

The steady-state expectation values of the atomic variables of interest can be calculated using the
expression for the steady-state solution of the master equation, i.e. Eq.~(\ref{master_solution}), and
by making use of the symmetrical collective states $|N,n,m \rangle$ corresponding to the SU(3) algebra
\cite{rev,ag_et,bog,mek2,mek}. The meaning of the symmetrical collective state $|N,n,m \rangle$ is
such that, $n$ atoms are considered to be in the bare state $|1\rangle$, $m - n$ in the state
$|2\rangle$, and $N-m$ in the bare state $|3\rangle$, where $N\geq n\geq0, ~N\geq m\geq n$.
Using the SU(3) eigenstate properties of the bare state atomic operators we can immediately
arrive at the expression for $Z$ (see, also \cite{Law}), i.e.,
\begin{equation}
 Z(\xi_1,\xi_3) = \frac{e^{-\xi_3N}}{1-e^{\xi_3}}[f(\xi_1-\xi_3)-e^{\xi_3(1+N)}f(\xi_1)].
\label{Z_eqn}
\end{equation}
Here,
$$ f(\xi) = \frac{1-e^{-\xi(1+N)}}{1-e^{-\xi}}$$.

Both the collective steady-state populations on the bare atomic states and their mutual correlations
can be calculated from the relations:
\begin{eqnarray}
\langle S^{k_{1}}_{11}S^{k_{2}}_{33}\rangle_{s} = (-1)^{k_{1}+k_{2}}Z^{-1}
\frac{\partial^{k_{1} + k_{2}}}{\partial \xi^{k_{1}}_{1}\partial \xi^{k_{2}}_{3}}Z(\xi_{1},\xi_{3}),
\label{SS}
\end{eqnarray}
with $\langle  S_{22} \rangle_s = N - \langle S_{11} \rangle_s - \langle S_{33} \rangle_s,
\hspace{0.1cm} \{k_{1},k_{2}=0,1,2 \cdots \}$.
In particular, for $N=1$ one obtains
\begin{eqnarray}
\langle S_{11}\rangle_{s} &=& \frac{\bar n_{1}\bar n_{2}}{(1+2\bar n_{1})(1+2\bar n_{2})-\bar n_{1}(1+\bar n_{2})},
\nonumber \\
\langle S_{22}\rangle_{s} &=& \frac{\bar n_{2}(1+\bar n_{1})}{(1+2\bar n_{1})(1+2\bar n_{2})-\bar n_{1}(1+\bar n_{2})},
\nonumber \\
\langle S_{33}\rangle_{s} &=& \frac{(1+\bar n_{1})(1+\bar n_{2})}{(1+2\bar n_{1})(1+2\bar n_{2})-\bar n_{1}
(1+\bar n_{2})}, \label{sol_1a}
\end{eqnarray}
and it can be observed that their corresponding ratios are in accordance with the equilibrium Boltzmann distribution.

In general, for any $N$, the population distribution of collection of atoms are given by:
\begin{eqnarray}
&{}&\langle S_{11}\rangle_{s} = \frac{\eta^{-N}_{2}}{Z(\xi_{1},\xi_{3})(\eta_{2}-1)}\biggl [\frac{(N+1)(\eta_{1}\eta_{2})^{N+1}-N(\eta_{1}\eta_{2})^{N+2} - \eta_{1}\eta_{2}}{(1-\eta_{1}\eta_{2})^{2}} \nonumber \\
&-&\eta^{N+1}_{2}\frac{(N+1)\eta^{N+1}_{1}-N\eta^{N+2}_{1}-\eta_{1}}{(1-\eta_{1})^{2}} \biggr ],   \nonumber \\
&{}&\langle S_{33}\rangle_{s} = \frac{\eta^{-N}_{2}}{Z(\xi_{1},\xi_{3})(\eta_{2}-1)^{2}} \nonumber\\
&\times& \biggl[ \frac{N  + (\eta_{1}\eta_{2})^{N+1}-\eta_{2}(\eta_{1}\eta_{2})^{N+2}-\eta_{1}\eta_{2}
(1-(N+2)\eta_{2}+N) -(N+1)\eta_{2}}{(1-\eta_{1}\eta_{2})^{2}} \nonumber \\
&+& \eta^{N+1}_{2}\frac{1-\eta^{N+1}_{1}}{1-\eta_{1}}\biggr], \label{sol_Na}
\end{eqnarray}
with $\langle S_{22}\rangle_{s} = N - \langle S_{11}\rangle_{s} - \langle S_{33}\rangle_{s}$ and
$\eta_{i}$ = $\bar n_{i}/[1+\bar n_{i}]$, $\{i \in 1,2 \}$. In this case, for larger $N$, the
collective interaction between the atoms drives the system into a thermal steady-state away from
a Boltzmann distribution \cite{mek}.

\begin{figure}[t]
\includegraphics[width = 7cm]{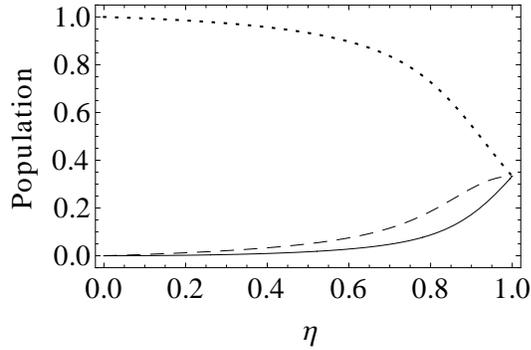}
\caption{\label{fig_1} The steady-state dependence of the scaled populations
$\langle S_{11}\rangle/N$ (solid line), $\langle S_{22}\rangle/N$ (dashed line)
and $\langle S_{33}\rangle/N$ (dotted line) as a function of $\eta \equiv \eta_{1} = \eta_{2}$.
Here $N=20$.}
\end{figure}

Now, we consider the following limiting cases of the applied incoherent field and the size of the system:

i) for a system with $\eta_{1}=0$, $\eta_{2} \not =0 $, we find that,
\begin{eqnarray*}
\langle S_{11}\rangle_{s} &=& 0 \\
\langle S_{33}\rangle_{s} &=& \frac{N -(N+1)\eta_{2}+\eta^{N+1}_{2}}{(1 - \eta_{2})(1 - \eta^{N+1}_{2})},
\end{eqnarray*}
i.e. we recovered the well-known results for a two-level ($|2\rangle \leftrightarrow |3\rangle$) atomic
sample \cite{has}.

ii) if $\{ \eta_{1} \not =0, \eta_{2} =0 \}$ or when there is no external incoherent pumping,
$\{ \eta_{1}=\eta_{2} =0 \}$, then $\langle S_{11}\rangle_{s} = 0$ and $\langle S_{33}\rangle_{s} = N$.
The system is entirely in the ground state.

iii) for a weak incoherent bath ($\eta_{i}<1$) and a large sample ($N \gg 1$) such that
$\{ (\eta_{1}\eta_{2})^{N}, \eta^{N}_{1(2)} \} \to 0$, we get,
\begin{eqnarray}
\langle S_{11}\rangle_{s} &=& \frac{\eta_{1}\eta_{2}}{1-\eta_{1}\eta_{2}}, \nonumber \\
\langle S_{22}\rangle_{s} &=& \frac{\eta_{2}}{1-\eta_{2}}, \nonumber \\
\langle S_{33}\rangle_{s} &=& N - \frac{\eta_{2}}{1-\eta_{2}} -\frac{\eta_{1}\eta_{2}}{1-\eta_{1}\eta_{2}},
\label{sol_NaL}
\end{eqnarray}

iv) on application of a strong incoherent field ($\eta_{i}=1$), we get,
\begin{eqnarray*}
\lim_{\eta_{1},\eta_{2}\to 1} \frac{\langle S_{11}\rangle_{s}}{N} =
\lim_{\eta_{1},\eta_{2}\to 1} \frac{\langle S_{22}\rangle_{s}}{N} =
\lim_{\eta_{1},\eta_{2}\to 1} \frac{\langle S_{33}\rangle_{s}}{N} =\frac{1}{3},
\end{eqnarray*}
i.e., the atomic levels are equally populated (see Fig.~\ref{fig_1}).

When cross-damping effects are considered \cite{int_et} then the corresponding expressions for
the population distributions can be obtained with the help of Eq.~(\ref{master_solution}) in
the limit $\eta_{1} = \eta_{2} \equiv \eta = \bar n/[1 + \bar n]$.
The mean value of the inversion operator $(S_{z}$ = $S_{11} - S_{33})$, can be evaluated using
the expression:
\begin{eqnarray}
\langle S^{k}_{z}\rangle_{s} &=& (-1)^{k}Z^{-1}\frac{\partial^{k}}{\partial \xi^{k}} Z(\xi),\hspace{0.7 cm }\{k=1,2,\cdots\} \label{Sz_k}
\end{eqnarray}
Hence,
\begin{eqnarray}
\langle S_{z}\rangle_{s} &=& - \frac{N+\eta^{N+1}+2\eta^{N+2}-(3+N)\eta^{3+2N}}{(1-\eta^{N+1})(1-\eta^{N+2})} \nonumber \\
&+& \frac{\eta(1+3\eta)}{1-\eta^{2}}.  \label{Sz}
\end{eqnarray}
We consider the following limiting cases of Eq.~(\ref{Sz}):

i) a system in a weak incoherent reservoir ($\eta \to 0$), has $\langle S_{z}\rangle_{s} = -N$;

ii) a fixed $\eta < 1$ and a large atomic sample, $N \gg 1$, such that
$\{ (\eta_{1}\eta_{2})^{N},\eta_{1(2)}^{N}\} \to 0 $, gives
$$
\langle S_{z}\rangle_{s} = -N + \frac{\eta(1+3\eta)}{1-\eta^{2}};
$$
and

iii) a strong heat bath ($\eta \to 1$) leads to saturation, i.e.
$$\lim_{\eta \to 1} \langle S_{z} \rangle_{s}/N =0.$$

Concluding this part, we emphasize that a weak incoherent reservoir leaves a large sample of
ladder emitters in their ground state, whereas equal distributions of populations among the
states is achieved on applying a strong incoherent field. The coherence terms, that appear
due to the quantum interference in the master equation do not affect the steady-state
behaviors of radiators on the bare atomic states. It can be realized here that the steady-state
inversion of atoms cannot be achieved in such a process, i.e.,
$\langle S_{11}\rangle_{s} \le \langle S_{22}\rangle_{s} \le \langle S_{33}\rangle_{s}$ or
$\langle S_{z}\rangle_{s}\leq0$ (see Fig.~\ref{fig_1}).
The above conclusions also apply for $V$-type emitters interacting with a thermal reservoir \cite{ag1}.

\section{First- and second-order correlation functions}
The degree of second-order coherence of collective fluorescent light emitted from the excited
atomic levels can be defined as \cite{qc1,qc2,wm},
\begin{eqnarray}
g^{(2)}(\vec r,t) = \frac{G^{(2)}(\vec r,t)}{[G^{(1)}(\vec r,t)]^{2}}, \label{g2_g}
\end{eqnarray}
with
\begin{eqnarray*}
G^{(1)}(\vec r,t) &=& \langle [E^{-}(\vec r,t)E^{+}(\vec r,t)]\rangle, \\
G^{(2)}(\vec r,t) &=& \langle :[ E^{-}(\vec r,t)E^{+}(\vec r,t)][E^{-}(\vec r,t)E^{+}(\vec r,t)]:\rangle,
\end{eqnarray*}
being the first- and the second-order correlation functions of the radiated EMF, respectively.
Here $E^{-}(\vec r,t)$ and $E^{+}(\vec r,t)$ are the positive and negative frequency parts of
the amplitude of the EMF operator $E$ at the space-point $\vec r$, and $: [\cdots] :$ means
normal ordering. In the far-zone limit of experimental interest, i.e.,
$r= |\vec r| \gg \{\lambda_{12},\lambda_{23}\}$, one can express the first- and the
second- order correlation functions via the collective atomic operators. Taking then the
long-time limit of Eq.~(\ref{g2_g}) and making use of Eq.~(\ref{master_solution}) the steady-state
coherence properties of the generated EMF will be investigated in the next subsections.

\subsection{Photon statistics of distinguishable photons}
Let us consider that the atomic transitions $|1\rangle\rightarrow|2\rangle$ and
$|2\rangle\rightarrow|3\rangle$ have dipole moments orthogonal to each other
$(\vec d_{12}\perp\vec d_{23})$. Then the emitted photons from the corresponding
transitions can be distinguished by their polarizations and frequencies and can
be detected by single-photon or two-photon detectors, respectively. For this case,
the normalized second-order coherence function can be defined as follows:
\begin{eqnarray}
g^{(2)}_{ij}(0) =
\frac{\langle J^+_iJ^+_jJ_jJ_i\rangle}{\langle J^+_iJ_i\rangle\langle J^+_jJ_j\rangle},
\hspace{0.5 cm }(i,j=1,2) \label{g2}
\end{eqnarray}
where, for brevity we have set $J_1 = S_{21}$ and $J_2 = S_{32}$, and $\langle J^+_iJ_i\rangle$
can be used to quantify the intensity of emitted light from the transition $i$. The quantity
$g^{(2)}_{ij}(0)$ can be interpreted as a measure for the probability for detecting one photon
emitted in transition $i$ and another photon emitted in transition $j$ simultaneously and its
value determines the nature of the emitted photons. $g^{(2)}_{ij}(0) < 1 $ characterizes
sub-Poissonian; $g^{(2)}_{ij}(0) > 1$, super-Poissonian; and $g^{(2)}_{ij}(0) = 1$, Poissonian
photon statistics of the emitted EMF. Anti-correlation or correlation of the emitted light occurs
when $g^{(2)}_{ij}(0)$ is smaller or larger than unity respectively. To evaluate these atomic
correlation functions, we can use the SU(3) eigenstate properties of the bare state atomic
operators \cite{rev,ag_et,bog,mek2,mek} and the exact steady-state solution,
i.e. the Eq.~(\ref{master_solution}).

Firstly, we evaluate the fluorescent steady-state intensities of light emitted on
$|1\rangle\rightarrow|2\rangle$ and $|2\rangle\rightarrow|3\rangle$, with help of the
following relations obtained from the master equation (\ref{master}):
\begin{eqnarray}
 G^{(1)}_1(0) &\propto& \langle S_{12}S_{21}\rangle_s \nonumber \\
 &=& \frac{\eta_1}{1-\eta_1}[N-\langle S_{33}\rangle_s - 2\langle S _{11}\rangle_s], \label{Int_1}
\end{eqnarray}
\begin{eqnarray}
 G^{(1)}_2(0) &\propto& \langle S_{23}S_{32}\rangle_s \nonumber \\
 &=& \frac{\eta_2}{1-\eta_2}[\langle S_{11}\rangle_s+2\langle S_{33}\rangle_s -N]. \label{Int_2}
\end{eqnarray}
Particularly, Eqs.~(\ref{Int_1}) and (\ref{Int_2}) were obtained from the steady-state form of the 
corresponding equations for $\langle S_{11}\rangle$ and $\langle S_{33}\rangle$ using also the 
commutation relations $[S_{12},S_{21}]=S_{11}-S_{22}$ and $[S_{23},S_{32}]=S_{22}-S_{33}$ as well as 
the relation $\langle S_{11}\rangle + \langle S_{22}\rangle + \langle S_{33}\rangle=N$. Consider the 
following limiting cases for $G^{(1)}_{1}(0)$ and $G^{(1)}_{2}(0)$:

i) if a weak incoherent field is applied to the sample, i.e., $\{\eta_{1},\eta_{2}\} \to 0$,
gives $G^{(1)}_{1}(0)$ = $G^{(1)}_{2}(0) =0$;

ii) large samples, $N \gg 1 $, with fixed $\{\eta_{1},\eta_{2}\} <1$, have
\begin{eqnarray}
G^{(1)}_{1}(0)&=& \frac{\eta_{1}}{1-\eta_{1}}\biggl [\frac{\eta_{2}}{1-\eta_{2}} -\frac{\eta_{1}\eta_{2}}{1-\eta_{1}\eta_{2}}\biggr], \label{G1_L1}
\end{eqnarray}
\begin{eqnarray}
G^{(1)}_{2}(0)&=& \frac{\eta_{2}}{1-\eta_{2}}\biggl [N -\frac{2\eta_{2}}{1-\eta_{2}} -\frac{\eta_{1}\eta_{2}}{1-\eta_{1}\eta_{2}}\biggr], \label{G1_L2}
\end{eqnarray}
and

iii) in the strong field limit $\{\eta_{1},\eta_{2}\} \to 1$ and fixed $N$, we find that,
\begin{eqnarray}
G^{(1)}_{1}(0) = G^{(1)}_{2}(0) = \frac{N}{12}(3+N). \label{G12_LL}
\end{eqnarray}
\begin{figure}[t]
\includegraphics[width = 7cm]{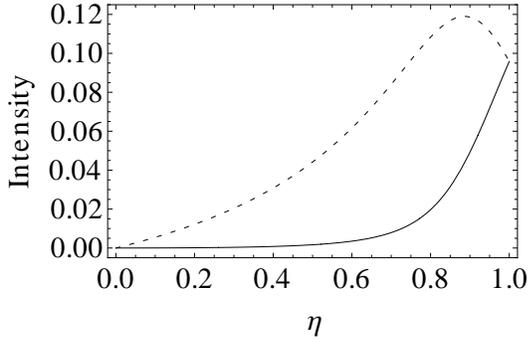}
\caption{\label{fig_2} The steady-state dependence of the scaled intensities $G^{(1)}_{1}(0)/N^2$
(solid line) and $G^{(1)}_{2}(0)/N^2$ (dashed line) as a function of
$\eta \equiv \eta_{1} = \eta_{2}$. Here $N=20$.}
\end{figure}

One can observe here that for larger atomic systems and moderate strengths of incoherent excitation
the first-order correlation function $G^{(1)}_1(0)$ does not depend on $N$ while $G^{(1)}_2(0)$
increases linearly with $N$, i.e., $G^{(1)}_2(0) \approx \bar n_2 N$. In the limit of intense
incoherent pumping the radiated fluorescence intensities in both atomic transitions scale as $N^2$,
similar to the superradiance phenomenon \cite{dik_et,rev}. Fig.~(\ref{fig_2}) depicts these intensities
as a function of the incoherent pumping strength. An interesting result here is that $G^{(1)}_{2}(0)/N^2$
shows a maximum for lower pumping intensities. From Eq.~(\ref{Int_2}), $G_2^{(1)}(0)$ can be written
as $\bar n_2[\langle S _{33}\rangle_s - \langle S_{22} \rangle_s] $. The value of $G_2^{(1)}(0)$
increases with the pumping parameter $\eta$. After a certain value, the value of $G_2^{(1)}(0)$ decreases
due to the rapidly decreasing nature of $[\langle S _{33}\rangle_s - \langle S_{22} \rangle_s]$, and
hence exhibits a maximum due to amplified spontaneous emission by the external incoherent excitation.

\begin{figure}[t]
\includegraphics[width = 7cm]{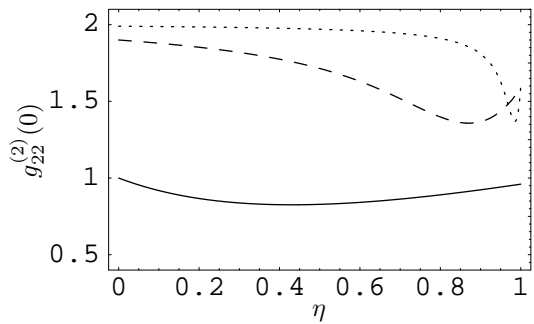}
\caption{\label{fig_3} The steady-state dependence of the second-order coherence function
$g^{(2)}_{22}(0)$ as function of $\eta \equiv \eta_{1} = \eta_{2}$. Here solid, dashed and
dotted curves are for $N = 2$, $20$, and $200$, respectively.}
\end{figure}

We now shall investigate the coherence properties of the light emitted on $|2\rangle \rightarrow|3\rangle$
atomic transition. For $N = 2$, the coherence factor $g^{(2)}_{22}(0)$ changes from unity (coherent light)
to values less than one (i.e., it exhibits sub-Poissonian photon statistics, see Fig.~\ref{fig_3}).
Hence, the emitted light possesses quantum features. For a moderately large atomic system, the fluorescent
field generated on this particular atomic transition has partial coherent properties because
$g^{(2)}_{22}(0) < 2$. The light-statistics of a large sample behaves as follows. For a weak bath
$(\eta < 1)$ it is incoherent since $\lim_{N\rightarrow\infty} g^{(2)}_{22}(0) = 2 $, showing the
super-Poissonian statistics of photons, while for an intense incoherent reservoir $(\eta = 1)$, it
is partially coherent, since
$$
\lim_{\eta\rightarrow1} g^{(2)}_{22}(0) = \frac{8(N-1)(N+4)}{5N(3+N)} \rightarrow \frac{8}{5}, ~\rm{when}~ N \gg 1.
$$

It should be noted here that the minimum for the coherence factor $g^{(2)}_{22}(0)$,
shown in Fig.~(\ref{fig_3}), occurs near the value of $\eta$ for which $G^{(1)}_2(0)$
is maximum, leading to the emission of quasi-coherent light. Therefore, there occurs an
enhancement of the multiparticle spontaneous emission corresponding to the maximum of
$G^{(1)}_2(0)$ and quasi-coherent light emission corresponding to the minimum of
$g^{(2)}_{22}(0)$ due to the surrounding incoherent reservoir and multi-level structure
of the emitters in the system. The incoherent pumping scheme developed here for orthogonal 
dipoles can be useful in higher frequency domains due to absence of good coherent sources. 
As can be seen from our results, one can obtain quasi-coherent light via incoherent pumping.

\subsection{Photon statistics of indistinguishable photons}
When decay interference effects are accounted, i.e., for near parallel dipoles
$(\vec d _{12} \parallel\vec d _{23})$, the second-order correlation function
can be represented as follows:
\begin{eqnarray}
g^{(2)}(0)= \frac{\langle (J^{+}_{1} + J^{+}_{2})^{2}(J_{1} +J_{2})^{2}\rangle}
{\langle (J^{+}_{1} + J^{+}_{2})(J_{1} + J_{2}) \rangle^{2}}.  \label{g2i}
\end{eqnarray}

It is emphasized here that due to the quantum decay interference the atomic transitions
are indistinguishable. The correlation function, in this case, is detected by a two-photon
detector. Eq.~(\ref{g2i}) contains off-diagonal terms that cannot be directly evaluated
with the solution obtained in Eq.~(\ref{master_solution}). However, we can represent the
correlation functions entering Eq.~(\ref{g2i}) via those atomic correlations that can be
evaluated with the steady-state solution in Eq.~(\ref{master_solution}). Therefore, using
the master equation, Eq.~(\ref{master}), we can show that:
\begin{eqnarray}
&{}&G^{(1)}(0)\propto \langle (J^{+}_{1} + J^{+}_{2})(J_{1} + J_{2})\rangle_{s}
= - \frac{\eta}{1-\eta}\langle S_{z}\rangle_{s}, \nonumber \\
&{}&G^{(2)}(0) \propto \langle (J^{+}_{1} + J^{+}_{2})^{2}(J_{1} + J_{2})^{2}\rangle_{s}
= \nonumber \\
&{}&\hspace{1.5cm}\frac{\eta^{2}}{(1-\eta )^{2}}
\bigg[ \frac{1 + \eta}{1-\eta}\langle S_{z}\rangle_{s} + 2\langle S^{2}_{z}\rangle_{s} \bigg],
\label{GG12}
\end{eqnarray}
and, thus, $g^{(2)}(0)$ can be written as,
\begin{eqnarray}
g^{(2)}(0)=\frac{[(1+\eta)/(1-\eta)]\langle S_{z}\rangle_{s} + 2\langle S^{2}_{z}\rangle_{s}}
{\langle S_{z}\rangle_{s} ^{2}},  \label{g2im}
\end{eqnarray}
with
\begin{eqnarray}
\langle S^{2}_{z}\rangle_{s} &=& \frac{a(\eta,N)\eta^{3+2N}-
b(\eta)\eta^{N+1}+c(\eta,N)}{(1-\eta)(1-\eta^{2})(1-\eta^{N+1})(1-\eta^{N+2})} \nonumber \\
&+& \frac{2\eta(1+3\eta)}{1-\eta^{2}}\langle S_{z}\rangle_{s}. \label{Sz2}
\end{eqnarray}
Here,
\begin{eqnarray*}
a(\eta,N)& = &(3+N)^{2}-(2+N)(4+N)\eta + N(6+N)\eta^{3} \\
&-& (1+N)(5+N)\eta^{2}, \\
b(\eta)& = &1+4\eta-8\eta^{3}-5\eta^{4}, \\
c(\eta,N)& = & N^{2}-(N^{2}-1)\eta-(N^{2}-4)\eta^{2}+(N^{2}-9)\eta^{3},
\end{eqnarray*}
and can be obtained from Eq.~(\ref{Sz_k}).
Setting $N=1$ in Eqs.~(\ref{Sz},\ref{Sz2}) we arrive at the corresponding expressions given in \cite{nar_et},
\begin{eqnarray*}
\langle S_{z}\rangle_{s} = \frac{\eta^{2}-1}{1 + \eta + \eta^{2}}, \hspace{0.3cm}
\langle S^{2}_{z}\rangle_{s} = \frac{\eta^{2}+1}{1 + \eta + \eta^{2}}
\end{eqnarray*}

The limiting cases for $G^{(1)}(0)$ in Eq.~(\ref{GG12}) are as follows:

i) for a weak incoherent field ($\eta \to 0$) and fixed $N$, it can be shown that $G^{(1)}(0)=0$ ;

ii) for a large sample ($N \gg 1$) and a fixed $ \eta < 1$, we find that
\begin{eqnarray}
G^{(1)}(0)= - \frac{\eta}{1-\eta}\bigg[-N + \frac{\eta(1+3\eta)}{1-\eta^{2}}\bigg] \approx \bar n N;
\label{cL1}
\end{eqnarray}
and

iii) for a strong incoherent field ($\eta \to 1$) and fixed $N$, we get
\begin{equation}
G^{(1)}(0) = \frac{N}{6}(3+N). \label{cL2}
\end{equation}
Thus, in the limit of intense pumping and large atomic samples the radiated fluorescence intensity
shows a quadratic dependence of $N$,  whereas  for moderate intensities it is a linear function of
$N$. In this case, for atoms emitting indistinguishable photons, the scaled intensity does not show
any maxima.
\begin{figure}[t]
\includegraphics[width = 7cm]{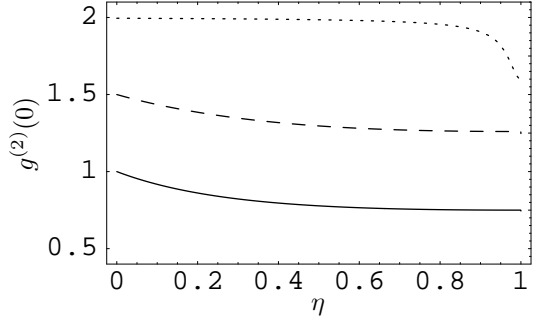}
\caption{\label{fig_4} Steady-state variance of the second-order coherence function $g^{(2)}(0)$
as a function of the parameter $\eta$. Solid, dashed and dotted curves are for N=1, 2, and 200,
respectively.}
\end{figure}

The values of the second-order correlation function for different limiting cases are as follows:

i) for  $N = 1$ and $\eta \not = 0$, we have,
$$
g^{(2)}(0) = 1-\frac{\eta}{(1+\eta)^{2}}.
$$
The emitted light shows sub-Poissonian photon statistics. Increasing the number of atoms leads to
partial coherence of the emitted electromagnetic field because $1< g^{(2)}(0)<2$  (see Fig.~\ref{fig_4}).

ii) in the limit of a weak incoherent field ($\eta \to 0$), we get
$$
g^{(2)}(0) = 2 - N^{-1},
$$

iii) in the limit of a strong incoherent bath ($\eta \to 1$), we obtain
\begin{equation}
g^{(2)}(0)= \frac{8N^{2}+24N-17}{5N(3+N)}. \label{g2_L}
\end{equation}
and

iv) for fixed $\eta <1$ and large samples, $N \gg 1$, one obtains
\begin{equation}
G^{(2)}(0) \approx 2 (\bar nN)^{2}, \label{ccL1}
\end{equation}
and, thus, $g^{(2)}(0)$ shows super-Poissonian photon statistics since in this case
$\lim_{N\to\infty}g^{(2)}(0) = 2$ [see Eqs.~(\ref{cL1},\ref{ccL1})]. Partial coherence
features occur for $\eta = 1$ and $N \gg 1$ because $\lim_{N\to\infty}g^{(2)}(0) = 8/5$
~[see Eq.~\ref{g2_L}].

The second-order correlation function for a three-level system with near parallel dipoles behaves
similar to that for a two-level sample \cite{has}. This can be seen also by introducing new atomic 
operators, i.e., $S^{+}=\sqrt{2}(S_{23} + S_{12})$, $S^{-}=\sqrt{2}(S_{21} + S_{32})$ and $S_{z}=S_{11}-S_{33}$ 
obeying the commutation relations for su(2) algebra: $[S^{+},S^{-}]=2S_{z}$ and $[S_{z},S^{\pm}]=\pm S^{\pm}$.
For equal decay rates, the master equation (\ref{master}) can be represented via new operators as follows:
\begin{eqnarray*}
\frac{d}{dt}\rho=-\frac{\gamma}{2}(1+\bar n)[S^{+},S^{-}\rho]-\frac{\gamma}{2}\bar n[S^{-},S^{+}\rho] + H.c.,
\end{eqnarray*}
and looks like the master equation describing two-level atoms \cite{has}. Hence, there is no amplified steady 
state spontaneous emission for such a system. 

\section{Summary}
The interaction of an ensemble of ladder-type emitters with an environmental incoherent reservoir
is investigated. The steady-state solution of the master equation and steady-state population
distributions for the system are obtained and it is shown that collective effects force the
system away from the Boltzmann-like thermodynamic equilibrium for systems with more than one atom.
Particularly, the ground-state emitters obey the Bose-Einstein statistics. We analyzed the photon 
statistics of the emitted light under different conditions. The emitted EMF in the case of one or 
two-atom sample emitting distinguishable photons, or a single atom emitting indistinguishable 
photons exhibit quantum properties. In case of atoms, emitting distinguishable photons, for 
larger samples, amplified steady-state spontaneous emission of quasi-coherent light occurs. 
Therefore, the investigated model can be useful in higher frequency domains as a source of 
quasi-coherent light. Furthermore, the first- and second-order coherence functions do not exhibit 
any critical behaviors, i.e., discontinuities or abrupt changes proper to phase transition phenomena.
Finally, the steady-state expectation values of any atomic variables of interest do not depend on 
spontaneous decay rates.

Rydberg atoms possessing almost equidistant energy levels and embedded inside a cavity with a low
quality factor are suitable candidates to test some of the results described here \cite{usp}. 
With suitable cavity parameters one can avoid the difficulties connected with the condition,
$\vec d _{12} \parallel \vec d_{23}$ \cite{sw}.

\section*{References}

\end{document}